\documentclass[aps,pra,10pt,reprint
]{revtex4-2}

\usepackage{graphicx}
\usepackage{dcolumn}
\usepackage{bm}

\usepackage{amsmath}
\usepackage{amssymb}
\usepackage{physics}
\usepackage{booktabs} 
\usepackage{array}

\usepackage{hyperref}
\begin{document}

\preprint{APS/123-QED}

\title{\textbf{Noise Propagation and Effective Channels in Encrypted Quantum Cloning} 
}%
\author{Yun-Fei Li}
\author{Fei-Hao Yang}%
 \author{Qing Chen}

 \author{Liu-Jun Wang}
 \thanks{Corresponding author. E-mail:ljwangq@ynu.edu.cn}
\affiliation{School of Physics and Astronomy and Yunnan Key Laboratory for Quantum Information,Yunnan University,Kunming 650500,People's Republic of China
}%


\date{\today}

\begin{abstract} 
We investigate noise propagation in encrypted quantum cloning by formulating the protocol in terms of effective recovery channels. For local single-qubit completely positive trace-preserving noise, a Pauli-basis expansion of the Kraus operators reveals a strong dependence on the location of the physical channel. Noise on an auxiliary signal qubit leaves the reduced recovered state unchanged, whereas noise on an auxiliary key qubit induces a Pauli-diagonal effective channel on the recovered qubit. For the core qubits, the error propagation additionally depends on the parity of the number of entangled pairs. In particular, for noise on the core key qubit with odd parity, distinct physical Pauli sectors merge into the same effective sector, converting isotropic depolarizing noise into an anisotropic recovery channel. We also relate Werner-type resource imperfections to local depolarizing noise and distinguish distributed product depolarization from genuine global depolarization. The results demonstrate that the effective recovery noise is governed not only by noise strength, but also by its physical location and by the structure of the encryption--decryption protocol.
\end{abstract}
\maketitle

\section{Introduction} 
The no-cloning theorem prohibits the perfect copying of an arbitrary unknown quantum state by any physical operation\cite{dieks1982communication,wootters1982single}. Its mixed-state counterpart, the no-broadcasting theorem, likewise rules out the perfect broadcasting of noncommuting states \cite{barnum1996noncommuting}. Nevertheless, approximate quantum cloning is possible and has been extensively investigated as a fundamental task in quantum information theory\cite{scarani2005quantum,buvzek1996quantum,gisin1997optimal,
bruss1998optimal,werner1998optimal,duan1998probabilistic,
fan2014quantum}.

Recently, Yamaguchi and Kempf introduced a different construction termed encrypted quantum cloning \cite{yamaguchi2026encrypted}. In this protocol, an unknown input qubit is coupled to several maximally entangled Bell pairs, and an encryption operation distributes the quantum information nonlocally among multiple signal and noise qubits. A subsequent decryption operation recovers the unknown state on a selected signal qubit. Under ideal conditions, the recovery is perfect while remaining consistent with the no-cloning theorem. The proposal has subsequently motivated studies of experimental realization, higher-dimensional generalizations, information-access structures, and connections with quantum secret sharing\cite{yamaguchi2026experimental,cearua2026cloning,
gianini2026encrypted,gianini2026full,lim2026encrypted,
gianini2026beyond}. These developments have primarily focused on the construction, implementation, and information structure of encrypted cloning; how physical noise acting at different locations of the protocol is transformed by the decryption procedure into an effective noise channel on the recovered qubit remains to be systematically characterized.

The ideal protocol, however, assumes maximally entangled Bell resources and noiseless operations. In realistic implementations, imperfections in state preparation, storage, and transmission can introduce decoherence and degrade the entangled resources \cite{bennett1996purification,bennett1996mixed,
horodecki2009quantum}. Werner-type states provide a simple benchmark for imperfect bipartite entanglement \cite{horodecki2009quantum,werner1989quantum,lee2000entanglement}, whereas general dynamical noise processes are naturally described by completely positive trace-preserving (CPTP) maps\cite{kraus1971general,choi1975completely,jamiolkowski1972linear,
ruskai2002analysis,fujiwara1999one}. This raises a central question: how does a physical noise channel acting on a particular subsystem propagate through the encrypted-cloning architecture, and what effective channel does it induce on the recovered qubit after decryption?

In this work, we develop a unified channel-based description of noise propagation in encrypted quantum cloning. We first use Werner-type resources as a physically transparent model of imperfect entanglement and show that imperfections of the core resource can lead to qualitatively different recovery channels depending on the parity of the number of entangled pairs. We then consider general local CPTP noise acting between encryption and decryption and analyze its propagation by expanding the corresponding Kraus operators in the Pauli basis. The resulting effective recovery channel depends crucially on the location of the physical noise. A trace-preserving channel acting solely on an auxiliary signal qubit leaves the reduced recovered state unchanged, whereas noise acting on a key qubit is converted into effective Pauli errors on the recovered qubit. For the core qubits, the error propagation additionally depends on the parity of the protocol; most notably, for noise acting on the core key qubit $N_1$ with odd $n$, distinct physical Pauli-error sectors merge into the same effective sector. We finally specialize the general framework to depolarizing noise, for which explicit recovery channels and fidelities can be obtained, and compare independently distributed depolarization with genuine global depolarization. These results show that the effective noise seen by the recovered qubit is determined not only by the noise strength, but also by its physical location and by the structure of the encryption–decryption protocol.

\begin{figure*}[t]
    \centering

    \includegraphics[
        width=1.0\textwidth
    ]{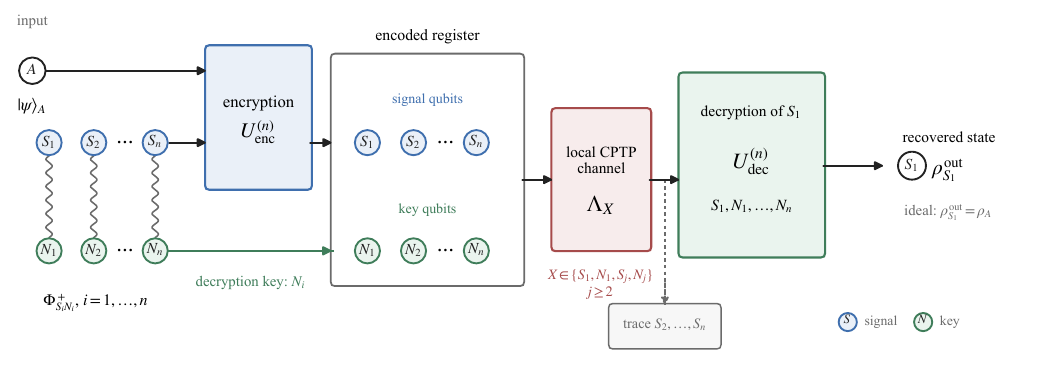}

    \caption{
    Schematic illustration of encrypted quantum cloning in the presence
    of quantum-channel noise.
    The input qubit $A$, initially in an unknown state $|\psi\rangle$,
    is encrypted together with the signal qubits
    $S_1,\ldots,S_n$, where each pair $(S_i,N_i)$ is initially
    prepared in a maximally entangled Bell state.
    The qubits $N_1,\ldots,N_n$ do not participate in encryption and
    collectively form the decryption key.
    After encryption, a local channel $\Lambda_X$ may act before the
    decryption operation.
    The auxiliary signal qubits $S_j$ $(j\geq2)$ do not participate in
    decryption and are subsequently traced out; consequently, as shown
    in Sec.~\ref{sec:general_channels}, any local trace-preserving channel acting solely on
    these qubits leaves the recovered state on $S_1$ unchanged.
    In contrast, the qubits $N_i$ participate explicitly in
    decryption, so local CPTP noise acting on them is propagated through
    the decryption operation into effective Pauli errors on the recovered
    qubit $S_1$.
    }

    \label{fig:noisy_protocol}
\end{figure*}

\section{Encrypted Cloning With Imperfect Entanglement Resources}
\label{sec:werner}

\subsection{Ideal protocol}
\label{subsec:ideal}
We first briefly review the ideal encrypted-cloning protocol of Ref.\cite{yamaguchi2026encrypted}. An unknown input qubit $A$ is prepared in the pure state
\begin{equation}
    \rho_A = |\psi\rangle\langle\psi|_A ,
\end{equation}
and is accompanied by $n$ maximally entangled Bell pairs
$(S_i,N_i)$,
\begin{equation}
    |\phi^+\rangle_{S_iN_i}
    =
    \frac{|00\rangle_{S_iN_i}+|11\rangle_{S_iN_i}}{\sqrt{2}} .
\end{equation}
where $n\ge1$.We denote
\begin{equation}
    \Phi^+_{S_iN_i}=|\phi^+\rangle\langle\phi^+|_{S_iN_i}.
\end{equation}
The initial state of the complete system is therefore
\begin{equation}
    \rho_0=\rho_A\otimes\bigotimes_{i=1}^{n}\Phi^+_{S_iN_i}.
\end{equation}

The qubits $S_i$ are referred to as signal qubits,
whereas the qubits $N_i$ are key qubits.
During encryption, the input qubit $A$ interacts with all signal
qubits through the unitary operator
\begin{equation}
    U_{\mathrm{enc}}^{(n)}=\frac{1}{2}\sum_{\mu=0}^{3}\alpha_{\mu}^{-1}\sigma_{\mu}^{(A)}
    \otimes
    \left(\bigotimes_{i=1}^{n}\sigma_{\mu}^{(S_i)}
    \right),
\end{equation}
where
\begin{equation}
    \alpha_0=1,\qquad \alpha_1=\alpha_3=i, \qquad\alpha_2=-i^{\,n+1},
\end{equation}
and $\{\sigma_\mu\}_{\mu=0}^{3}$ denotes the identity and
the three Pauli operators.

The encryption unitary $U_{\rm enc}^{(n)}$ acts trivially on all key qubits $N_i$. Nevertheless, the encoded information is not confined to the signal subsystem. Because each 
$S_i$ is initially entangled with $N_i$, the encryption operation modifies the joint correlations among $A$, the signal qubits, and the key qubits. Consequently, although the $N_i$ are not directly acted upon during encryption, their correlations with the signal qubits are essential for the subsequent recovery of the input state.

To recover the state on the target signal qubit $S_1$, one apply the
decryption unitary 
\begin{equation}
    U_{\mathrm{dec}}^{(n)} = \sum_{\mu=0}^{3} \alpha_\mu
    \left( |\phi_\mu\rangle\langle\phi_\mu| \right)_{S_1N_1}
    \otimes
    \left(\bigotimes_{i=2}^{n}\sigma_{\mu}^{(N_i)T}
    \right),
\end{equation}
where
\begin{equation}
    |\phi_\mu\rangle_{S_iN_i} =\sigma_{\mu}^{(S_i)}|\phi^+\rangle_{S_iN_i}.
\end{equation}

Thus, recovery of $S_1$ involves $S_1$ and all key qubits $N_1,\ldots,N_n$, whereas the auxiliary signal qubits $S_j$ with $j\ge2$ do not participate in the decryption unitary.

Under ideal conditions, encryption followed by decryption perfectly transfers the input state to the target signal qubit:
\begin{equation}
    \operatorname{Tr}_{\overline{S_1}}
    \left[ U_{\mathrm{dec}}^{(n)} U_{\mathrm{enc}}^{(n)}\rho_0U_{\mathrm{enc}}^{(n)\dagger}U_{\mathrm{dec}}^{(n)\dagger}
    \right]
    =
    |\psi\rangle\langle\psi|_{S_1}
    \equiv
    \rho_A^{(S_1)} .
\end{equation}
Here $\overline{S_1}=AN_1S_2N_2\dots S_nN_n$ denote the complement of $S_1$,Hence, the unknown input state is recovered exactly on $S_1$.

The complete encryption–noise–decryption architecture considered in this work is summarized in Fig.\ref{fig:noisy_protocol}. The figure also highlights the structural distinction that will be central to the noise analysis below: the auxiliary signal qubits $S_j$ $(j\ge2$ are excluded from the decryption unitary and subsequently traced out, whereas the key qubits $N_i$ participate directly in the recovery operation.

\subsection{Werner-type imperfect entanglement resources}
\label{subsec:Werner resource}
The ideal protocol assumes maximally entangled resource pairs. In practice, imperfections in state preparation, storage, or transmission can degrade these resources into mixed states. To isolate the effect of imperfect entanglement in the pair directly involved in the Bell-projector part of the decryption unitary, we first replace the core pair $S_1N_1$ by the Werner-type state\cite{werner1989quantum}
\begin{equation}
\label{eq:S1N1 werner}
    W_{p_1}^{(S_1N_1)}=p_1\Phi^+_{S_1N_1}+\frac{1-p_1}{4}I_{S_1N_1},
    \qquad
    0\leq p_1\leq1 .
\end{equation}
Here $p_1$ quantifies the quality of the resource: $p_1=1$ corresponds to the ideal Bell pair, whereas $p_1=0$ gives the maximally mixed two-qubit state.

When only the core pair $S_1N_1$ is imperfect and all remaining resource pairs are ideal, the initial state becomes
\begin{equation}
    \rho_0^{W}=\rho_A\otimes W_{p_1}^{(S_1N_1)}\otimes\bigotimes_{i=2}^{n}\Phi^+_{S_iN_i}.
\end{equation}
Since the encryption–decryption transformation and the final partial trace are linear, the Bell-state and maximally mixed contributions in $W_{p_1}^{(S_1N_1)}$ can be propagated separately.

To expose the origin of the parity dependence, consider the maximally mixed contribution. Using the Bell basis generated by local Pauli operations,
\begin{equation}
    I_{S_1N_1} = \sum_{\mu=0}^{3} \sigma_\mu^{(S_1)} \Phi^+_{S_1N_1}
    \sigma_\mu^{(S_1)} .
\end{equation}

Hence, $I_{S_1N_1}/4$ is an equal mixture of the four Pauli-transformed Bell states. Propagating these four Bell sectors through the encryption–decryption protocol shows that their contributions to the recovered state depend on the parity of the total number $n$ of entangled pairs; the detailed derivation is given in the Supplemental Material.

For even $n$, the maximally mixed component of the core resource is mapped to the maximally mixed state of the recovered qubit. The output therefore takes the form
\begin{equation}
\label{eq:even n werner}
    \rho_{S_1}^{\mathrm{out}} = p_1\rho_A^{(S_1)} + (1-p_1)\frac{I_{S_1}}{2},
    \qquad n~\mathrm{even}.
\end{equation}
Thus, core-resource imperfection induces an isotropic depolarizing form on the recovered qubit. The ideal result is recovered for $p_1=1$, while $p_1=0$ yields $I_{S_1}/2$.

For odd $n$, however, the maximally mixed part of the core resource does not produce a maximally mixed recovered qubit. Instead,
\begin{equation}
\label{eq:odd n werner}
    \rho_{S_1}^{\mathrm{out}}=\frac{1+p_1}{2}\rho_A^{(S_1)}+\frac{1-p_1}{2}\sigma_2^{(S_1)}\rho_A^{(S_1)}\sigma_2^{(S_1)},
    \qquad n~\mathrm{odd}.
\end{equation}
The resource imperfection therefore induces an anisotropic effective Pauli error rather than isotropic mixing. In particular, Eq.(\ref{eq:odd n werner}) has the form of a $Y$-type Pauli channel on the recovered qubit, with error probability $(1-p_1)/2$.

Consequently, the $y$ axis of the Bloch sphere is a protected direction of this effective channel. If the input state is an eigenstate of $\sigma_2$,
\begin{equation}
    \sigma_2\rho_A\sigma_2=\rho_A,
\end{equation}
and Eq.(\ref{eq:odd n werner}) reduces to
\begin{equation}
    \rho_{S_1}^{\mathrm{out}}
    =
    \rho_A^{(S_1)}.
\end{equation}
Such states can therefore remain perfectly recoverable even in the
presence of core-resource imperfection.

Equations (\ref{eq:even n werner}) and (\ref{eq:odd n werner}) show that the same form of core-resource imperfection is converted into qualitatively different effective recovery noise depending on the parity of $n$: even $n$ produces isotropic mixing, whereas odd $n$ selects a preferred Pauli axis.

The analysis can be extended to several imperfect resource pairs, as shown in the Supplemental Material. However, treating each resource configuration separately quickly becomes cumbersome and obscures the common structure of the resulting errors. A more unified description is obtained by viewing resource degradation, whenever appropriate, as arising from an underlying quantum channel. This motivates the channel-based formulation introduced in the following subsection.

\subsection{From resource imperfections to channel noise}
\label{subsec:Werner resource to channel noise}
The Werner-type description used above treats the imperfection as a property of the initially prepared entangled resource, without specifying the dynamical process responsible for its degradation. For the particular family considered in Eq.(\ref{eq:S1N1 werner}), however, the same resource state can be generated dynamically by applying a local depolarizing channel to one subsystem of an initially maximally entangled Bell pair.

Consider the single-qubit depolarizing channel
\begin{equation}
    \mathcal{D}_p(\rho) = p\rho + (1-p)\frac{I}{2},
    \qquad
    0\leq p\leq1 .
\end{equation}
When $\mathcal D_p$ acts on one member of an ideal Bell pair,
\begin{equation}
    \left(\mathcal{I}\otimes\mathcal{D}_p\right)
    \left(\Phi^+\right)=p\Phi^++\frac{1-p}{4}I.
\end{equation}
This is precisely the Werner-type resource introduced in Eq.(\ref{eq:S1N1 werner}). Hence, for this class of imperfections, an initially noisy entangled resource may equivalently be interpreted as an ideal Bell pair that has undergone a local depolarizing channel. More generally, such channel--state connections are closely related to teleportation-based channel simulation, in which teleportation-covariant channels, including Pauli channels, can be simulated using their corresponding Choi resource states\cite{pirandola2017fundamental}.
The parameters $p_i$ appearing in the Werner-resource description
can therefore be regarded as local channel-quality parameters.

The equivalence becomes especially useful for the key qubits $N_i$. Since $U_{\rm enc}^{(n)}$ acts trivially on every $N_i$, a local depolarizing channel on $N_i$ commutes with the encryption channel:
\begin{equation}
    \mathcal{D}_{p_i}^{(N_i)}\circ{U}_{\mathrm{enc}}^{(n)}=
    {U}_{\mathrm{enc}}^{(n)}\circ{D}_{p_i}^{(N_i)},
\end{equation}
where ${U}_{\mathrm{enc}}^{(n)}(\rho)=U_{\mathrm{enc}}^{(n)}\rho U_{\mathrm{enc}}^{(n)\dagger}.$ Therefore, within the present protocol, depolarizing noise acting on a key qubit $N_i$ after encryption can be moved to the resource-preparation stage without changing the state entering the decryption operation.

This time-shifting argument does not hold in general. If a channel $\Lambda_X$ acts on a subsystem $X$ on which $U_{\rm enc}^{(n)}$ acts nontrivially, there is in general no reason for the two maps to commute:
\begin{equation}
    \Lambda_X\circ{U}_{\mathrm{enc}}^{(n)}\neq
    {U}_{\mathrm{enc}}^{(n)}\circ\Lambda_X .
\end{equation}
Consequently, an imperfect initial resource and a dynamical
post-encryption noise process cannot in general be regarded as
physically equivalent.

The Werner-resource analysis thus provides a useful bridge between static resource imperfections and dynamical channel noise. It also makes clear, however, that a resource-state description alone does not capture the location and timing of a general physical noise process. We therefore turn in Sec.\ref{sec:general_channels} to quantum channels acting directly on the encrypted state between the encryption and decryption stages, and determine how the resulting effective recovery channel depends on the physical location of the noise.

\section{Encrypted Quantum Cloning under General Quantum Channels}
\label{sec:general_channels}

\subsection{General framework}
\label{sec:general_framework}

We now consider a physical noise process acting on the encrypted state between the encryption and decryption stages. Our goal is to determine the effective recovery channel induced on the target qubit $S_1$ by a noise channel acting on a specified physical subsystem $X$. After encryption, the joint state is
\begin{equation}
    \rho_{\mathrm{enc}}=U_{\mathrm{enc}}^{(n)}\rho_0U_{\mathrm{enc}}^{(n)\dagger}.
    \label{eq:rho_enc_general}
\end{equation}

Let a physical subsystem $X$ undergo a completely positive trace-preserving (CPTP) map $\Lambda_X$\cite{kraus1971general,choi1975completely,jamiolkowski1972linear,ruskai2002analysis},
\begin{equation}
    \Lambda_X(\rho)=\sum_rE_r^{(X)}\rho E_r^{(X)\dagger},\qquad\sum_rE_r^{(X)\dagger}E_r^{(X)}=I_X.
    \label{eq:general_channel}
\end{equation}

After applying the ideal decryption unitary and tracing out all subsystems except $S_1$, the recovered state is
\begin{equation}
    \rho_{S_1}^{\mathrm{out}}
    =
    \operatorname{Tr}_{\overline{S_1}}
    \left[
        U_{\mathrm{dec}}^{(n)}
        \Lambda_X(\rho_{\mathrm{enc}})
        U_{\mathrm{dec}}^{(n)\dagger}
    \right]=\mathcal{E}_X^{\mathrm{eff}}(\rho_A),
    \label{eq:general_output}
\end{equation}
where $\mathcal E_X^{\rm eff}:\mathcal{B}(\mathcal{H}_A)\to\mathcal{B}(\mathcal{\mathcal{H}}_{S_1})$ denotes the effective recovery map associated with physical noise acting at location $X$,and $\mathcal{B}(\mathcal{H})$ represents the set of linear operators defined on the Hilbert space $\mathcal{H}$.

A convenient starting point is the factorized form of the ideal globally decrypted state,
\begin{equation}
    U_{\mathrm{dec}}^{(n)}\rho_{\mathrm{enc}}U_{\mathrm{dec}}^{(n)\dagger}=\rho_A^{(S_1)}\otimes\Omega,
\end{equation}
where$\Omega =\Phi_{AN_1}^{+}\otimes\bigotimes_{i=2}^{n} \Phi_{S_iN_i}^{+}.$ Defining the propagated Kraus operators
\begin{equation}
\label{eq:Kraus expansion}
    \widetilde{E}_r^{(X)}=U_{\mathrm{dec}}^{(n)}E_r^{(X)}
    U_{\mathrm{dec}}^{(n)\dagger},
\end{equation}

These operators describe how a physical error occurring before decryption is represented in the ideally decrypted frame. Using Eq.(\ref{eq:Kraus expansion}), Eq.(\ref{eq:general_output}) becomes
\begin{equation}
    \rho_{S_1}^{\mathrm{out}}
    =
    \sum_r
    \operatorname{Tr}_{\overline{S_1}}
    \left[
        \widetilde{E}_r^{(X)}
        (\rho_A^{(S_1)}\otimes\Omega)
        \widetilde{E}_r^{(X)\dagger}
    \right].
    \label{eq:master_formula}
\end{equation}
Equation (\ref{eq:master_formula}) is the basic noise-propagation formula used below.

For a local single-qubit channel, we expand the Kraus
operators in the Pauli basis \cite{ruskai2002analysis,fujiwara1999one}.
\begin{equation}
    E_r^{(X)}=\sum_{a=0}^{3}C_{ra}^{(X)}\sigma_a^{(X)}.
    \label{eq:pauli_expansion}
\end{equation}

Equivalently, the physical channel can be written in the Pauli process-matrix form
\begin{equation*}
    \Lambda_X(\rho)=\sum_{a,b=0}^3\xi^{(X)}_{ab}\sigma_a^{(X)}\rho\sigma_b^{(X)}
\end{equation*}
where
\begin{equation*}
    \xi^{(X)}_{ab}=\sum_rC_{ra}^{(X)}C_{rb}^{(X)*}
\end{equation*}
We denote the diagonal Pauli weights by
\begin{equation*}
    q_a\equiv\xi^{(X)}_{aa}=\sum_r|C_{ra}^{(X)}|^2,\quad\sum_aq_a=1
\end{equation*}

The off-diagonal elements \(\xi_{ab}^{(X)}\) with \(a\neq b\) represent coherences between different Pauli components. For the local single-qubit noise configurations considered below, the decryption structure and the final partial trace eliminate these cross terms, so that the recovered qubit is governed by location-dependent combinations of the diagonal weights \(q_a\).

The local noise-propagation problem is therefore reduced to determining how the four Pauli components are transformed by \(U_{\rm dec}^{(n)}\) and which of the resulting operators survive the final partial trace. As shown below, this transformation depends strongly on the physical location of the noise and, for the core qubits, on the parity of \(n\).

\subsection{Noise on auxiliary signal and key qubits}
\label{sec:auxiliary_noise}
The auxiliary pairs $(S_j,N_j)$, with $j\ge2$, exhibit a fundamental asymmetry between their signal and key qubits. The signal qubit $S_j$ does not participate in the decryption unitary and is ultimately traced out, whereas the corresponding key qubit $N_j$ enters explicitly into $U_{\rm dec}^{(n)}$. As a result, the two locations lead to qualitatively different effective recovery channels.

Consider first a local CPTP channel acting on an auxiliary signal qubit \(S_j\), \(j\ge2\). Since \(S_j\) does not participate in the decryption unitary,
\begin{equation*}
    \Lambda_{S_j}\circ\mathcal{U}_{\rm dec}^{(n)}=\mathcal{U}_{\rm dec}^{(n)}\circ\Lambda_{S_j}
\end{equation*}
Moreover, \(S_j\) belongs to the subsystem traced out at the end of the protocol. For any trace-preserving map,
\begin{equation*}
    Tr_{S_j}[(\Lambda_{S_j}\otimes\mathcal{I}(\rho)]=Tr_{S_j}(\rho)
\end{equation*}
Therefore(detailed computation see in the supplement material),
\begin{equation}
    \rho_{S_1}^{\mathrm{out}}
    =\rho_A^{(S_1)},
    \qquad
    X=S_j,\quad j\geq2.
    \label{eq:Sj_no_effect}
\end{equation}
Equivalently, \(\mathcal E_{S_j}^{\rm eff}=\mathcal I\). This statement concerns the reduced recovered state on \(S_1\); the channel may still modify the global state and correlations involving \(S_j\).

The situation is different for an auxiliary key qubit $N_j$, since \(N_j\) participates explicitly in the decryption unitary. Using the Bell-state identity
\begin{equation*}
    (\sigma_a\otimes I)|\phi^+\rangle=(I\otimes\sigma_a^T)|\phi^+\rangle
\end{equation*}
it is convenient to expand the Kraus operators in the transposed Pauli basis,
\begin{equation}
    E_r^{(N_j)} = \sum_{a=0}^{3}
    C_{ra}
    \sigma_a^{(N_j)\mathsf T},
\end{equation}
one finds that each Pauli component is propagated onto the recovered
qubit $S_1$. After tracing out the remaining systems(detailed computation see in the supplement material),
\begin{equation}
    \rho_{S_1}^{\mathrm{out}} =\sum_{a=0}^{3}q_a\sigma_a^{(S_1)}\rho_A^{(S_1)}\sigma_a^{(S_1)},
    \qquad X=N_j,\quad j\geq2,
    \label{eq:Nj_effective_pauli}
\end{equation}

where
\begin{equation*}
    q_a=\sum_r|C_{ra}|^2=\xi_{aa},\quad\sum_aq_a=1
\end{equation*}
Thus, an arbitrary local single-qubit CPTP channel acting on \(N_j\) induces a Pauli-diagonal effective channel on the recovered qubit.

Equivalently, define the Choi state
\begin{equation}
    \omega_{S_jN_j}=
    (I_{S_j}\otimes\Lambda_{N_j})(\Phi_{S_jN_j}^{+}),
\end{equation}
The effective Pauli weights are precisely its Bell-basis populations,
\begin{equation}
    q_a=\langle\phi_a|
    \omega_{S_jN_j}
    |\phi_a\rangle.
    \label{eq:qa_bell}
\end{equation}
Hence, the reduced recovery channel retains only the Bell-diagonal part of the Choi state of the physical noise.

For the depolarizing channel,
\begin{equation}
    E_0=\sqrt{\frac{1+3p}{4}}I,\quad E_a=\sqrt{\frac{1-p}{4}}\sigma_a,\;a=1,2,3
    \label{kraus ofdepolarize}
\end{equation}
giving
\begin{equation}
    q_0=\frac{1+3p}{4},\quad q_1=q_2=q_3=\frac{1-p}{4}
\end{equation}
And for bit-flip channel\cite{nielsen2010quantum}
\begin{equation*}
    \mathcal{B}_f(\rho)=(1-f)\rho+f\sigma_1\rho\sigma_1
    \label{bit-flip}
\end{equation*}
one has
\begin{equation}
    q_0=1-f,\quad q_1=f,\quad q_2=q_3=0
\end{equation}
 Finally,for a generalized amplitude damping channel\cite{srikanth2008squeezed,nielsen2010quantum},its Kraus representation is
 \begin{equation*}
     \begin{split}
        & E_0=\sqrt{1-p}
         \begin{pmatrix}
             1&0\\
             0&\sqrt{1-\gamma}
         \end{pmatrix},\qquad
         E_1=\sqrt{1-p}
         \begin{pmatrix}
           0&\sqrt{\gamma}\\
           0&0
         \end{pmatrix}\\
         &E_2=\sqrt{p}
         \begin{pmatrix}
            \sqrt{1-\gamma}&0\\
            0&1
         \end{pmatrix},\qquad 
         E_3=\sqrt{p}
         \begin{pmatrix}
             0&0\\
             \sqrt{\gamma}&0
         \end{pmatrix}
     \end{split}
     \label{GAD Kraus}
 \end{equation*}
 with
 \begin{equation}
     q_0=\frac{(1+\sqrt{1-\gamma})^2}{4},\; q_1=q_2=\frac{\gamma}{4},\; q_3=\frac{(1-\sqrt{1-\gamma})^2}{4}
 \end{equation}
 
 Remarkably, these effective weights are independent of the bath-population parameter of the generalized amplitude-damping channel. This does not imply that the underlying physical channels are identical; rather, the differences not encoded in their Bell-basis populations are removed in the reduced recovery channel.

 Hence, local noise on an auxiliary key qubit \(N_j\) is converted by the encrypted-cloning recovery procedure into an effective Pauli channel on \(S_1\), whose probabilities are given by the Bell-basis populations of the Choi state of the physical channel. This relation is closely connected to earlier results linking noisy entangled resources and effective teleportation channels\cite{horodecki1999general,bowen2001teleportation,ziman2005concurrence,bennett1993teleporting,verstraete2003optimal}. The present setting differs in that the effective channel arises from the location-dependent propagation of physical errors through the encrypted-cloning decryption structure.

\subsection{Core-qubit noise and parity dependence}
\label{sec:core qubits}

Noise acting on the core pair $S_1N_1$ is qualitatively different from noise on the auxiliary pairs. 
Both $S_1$ and $N_1$ enter directly into the Bell-projector sector of the decryption unitary, so propagating a local Pauli component through $U_{\rm dec}^{(n)}$ generally produces a multi-qubit Pauli string. 
After the complementary subsystems are traced out, the induced Pauli action on the recovered qubit depends on the parity of $n$. 
The explicit operator identities are derived in the Supplemental Material.

\subsubsection{Noise on \texorpdfstring{$S_1$}{S\_1}}

Consider first a local CPTP channel acting on the core signal qubit $S_1$. 
As in Sec.\ref{sec:general_framework}, let
\[
q_a=\xi_{aa}
\]
denote the diagonal Pauli weights of the physical channel. 
Propagating the Pauli components through the decryption unitary and tracing the complementary subsystems gives the following effective Pauli action on the recovered qubit:
\begin{equation}
\begin{array}{c|ccc}
 & \sigma_1 & \sigma_2 & \sigma_3\\
\hline
n\ {\rm even}
& \sigma_3 & \sigma_1 & \sigma_2\\
n\ {\rm odd}
& \sigma_3 & \sigma_2 & \sigma_1
\end{array}.
\label{eq:S1_effective_mapping}
\end{equation}
Here the table refers only to the induced action on $S_1$ after the remaining subsystems are traced out; the full conjugated operators are multi-qubit Pauli strings, as shown in the Supplemental Material.

For even $n$, the recovered state is
\begin{align}
\rho_{S_1}^{\rm out}
={}&q_0\rho_A^{(S_1)}
+q_1\sigma_3^{(S_1)}
\rho_A^{(S_1)}
\sigma_3^{(S_1)}
\nonumber\\
&+q_2\sigma_1^{(S_1)}
\rho_A^{(S_1)}
\sigma_1^{(S_1)}
+q_3\sigma_2^{(S_1)}
\rho_A^{(S_1)}
\sigma_2^{(S_1)},
\quad n\ {\rm even}.
\label{eq:S1_even}
\end{align}
For odd $n$, one instead obtains
\begin{align}
\rho_{S_1}^{\rm out}
={}&q_0\rho_A^{(S_1)}+q_1\sigma_3^{(S_1)}
\rho_A^{(S_1)}
\sigma_3^{(S_1)}
\nonumber\\
&+q_2\sigma_2^{(S_1)}\rho_A^{(S_1)}\sigma_2^{(S_1)}+q_3\sigma_1^{(S_1)}
\rho_A^{(S_1)}\sigma_1^{(S_1)},
\qquad n\ {\rm odd}.
\label{eq:S1_odd}
\end{align}

In both cases, the identity sector remains unchanged and the four physical Pauli sectors remain distinct. 
Parity therefore changes only the permutation assigned to the three nontrivial Pauli weights. 
In particular, when
\begin{equation}
q_1=q_2=q_3,
\label{eq:isotropic_condition}
\end{equation}
the permutation becomes irrelevant and the recovered state is parity independent. 
This is the case for isotropic depolarizing noise discussed in Sec.\ref{sec:depolarizing noise}.

\subsubsection{Noise on \texorpdfstring{$N_1$}{N\_1}}

A stronger parity effect appears when the physical channel acts on the core key qubit $N_1$. 
In contrast to the $S_1$ case, odd parity changes not only the assignment of Pauli errors but also the number of distinct effective error sectors.

For even $n$, the effective Pauli sectors remain one-to-one:
\begin{equation}
I\longrightarrow I,\qquad
\sigma_1\longrightarrow\sigma_2,\qquad
\sigma_2\longrightarrow\sigma_3,\qquad
\sigma_3\longrightarrow\sigma_1.
\label{eq:N1_even_mapping}
\end{equation}
Here, as in Eq.~\eqref{eq:S1_effective_mapping}, the arrows denote the induced action on the recovered qubit after the complementary subsystems have been traced out. 
The corresponding recovered state is
\begin{align}
\rho_{S_1}^{\rm out}
={}&q_0\rho_A^{(S_1)}+q_1\sigma_2^{(S_1)}\rho_A^{(S_1)}\sigma_2^{(S_1)}
\nonumber\\
&+q_2\sigma_3^{(S_1)}\rho_A^{(S_1)}\sigma_3^{(S_1)}
+q_3\sigma_1^{(S_1)}\rho_A^{(S_1)}\sigma_1^{(S_1)},\qquad n\ {\rm even}.
\label{eq:N1_even}
\end{align}
Thus, for even $n$, the physical Pauli weights are merely reassigned among the effective sectors.

For odd $n$, however, the effective mapping is no longer one-to-one:
\begin{equation}
I,\sigma_2\longrightarrow I,
\qquad
\sigma_1,\sigma_3\longrightarrow\sigma_2.
\label{eq:N1_odd_mapping}
\end{equation}
Consequently, the recovered state becomes
\begin{equation}
\rho_{S_1}^{\rm out}=(q_0+q_2)\rho_A^{(S_1)}+(q_1+q_3)
\sigma_2^{(S_1)}\rho_A^{(S_1)}
\sigma_2^{(S_1)},
\qquad n\ {\rm odd}.
\label{eq:N1_odd}
\end{equation}

Equation~\eqref{eq:N1_odd} exhibits a genuine merging of Pauli sectors rather than a simple permutation. 
The physical $I$ and $\sigma_2$ components jointly contribute to the effective identity sector with total weight $q_0+q_2$, whereas the physical $\sigma_1$ and $\sigma_3$ components jointly contribute to a single effective $\sigma_2$ sector with weight $q_1+q_3$. 
Thus, information distinguishing the two physical errors within each pair is absent from the reduced recovery channel. 
Equivalently, for odd $n$ the effective channel may be written as
\begin{equation}
\mathcal{E}_{N_1}^{\rm eff}(\rho)=(q_0+q_2)\rho+(q_1+q_3)\sigma_2\rho\sigma_2.
\label{eq:N1_odd_effective_channel}
\end{equation}

The contrast between $S_1$ and $N_1$ is therefore structural. 
For noise on $S_1$, changing the parity of $n$ only permutes the nontrivial Pauli sectors. 
For noise on $N_1$, odd $n$ instead produces a many-to-one coarse graining of the physical Pauli errors. 
Consequently, even an isotropic physical Pauli distribution can be converted into an anisotropic effective recovery channel. 
This effect becomes particularly transparent for depolarizing noise in Sec.\ref{sec:depolarizing noise}.

\subsection{Correlated noise on the decryption subsystem}
\label{sec:correlated_noise}
The preceding analysis concerns local single-qubit channels. More generally, several qubits participating in the recovery operation may couple to a common environment and undergo correlated noise. Correlated multiqubit channels and quantum channels with memory have been extensively studied in quantum information theory\cite{macchiavello2002entanglement,yeo2003quantum,daffer2004depolarizing,kretschmann2005quantum,caruso2010teleportation,RevModPhys.86.1203}. Here we consider correlated noise acting on the subsystem directly involved in decryption.

We denote by
\begin{equation}
    B = S_1N_1N_2\cdots N_n
\end{equation}
the subsystem on which the decryption unitary acts nontrivially, and consider a CPTP map
\begin{equation}
    \Lambda_B(\rho)=\sum_rF_r^{(B)}\rho F_r^{(B)\dagger},\quad\sum_r F^{(B)\dagger}_rF_r^{(B)}=I_B
\end{equation}

Each Kraus operator can be expanded in the multiqubit Pauli basis,
\begin{equation}
    F_r^{(B)}=\sum_{\mathcal E}C_{r;\mathcal E}P_{\mathcal E},
    \label{eq:correlated_pauli_expansion}
\end{equation}
where $\mathcal{E}=(e_{S_1},e_{N_1},\ldots,e_{N_n})$ labels a Pauli string, with $e_X\in\{0,1,2,3\}$, and
\begin{equation*}
    P_{\mathcal{E}}=\sigma_{eS_1}^{(S_1)} \otimes \sigma_{eN_1}^{(N_1)} \otimes \cdots \otimes \sigma_{eN_n}^{(N_n)}.
\end{equation*}
Defining the corresponding propagated Pauli strings
\begin{equation}
\label{eq:propagated pauli string}
    W_{\mathcal E} = U_{\mathrm{dec}}^{(n)} P_{\mathcal E} U_{\mathrm{dec}}^{(n)\dagger},
\end{equation}
the recovered state can be written as
\begin{equation}
\begin{aligned}
    \rho_{S_1}^{\mathrm{out}}=\sum_r\sum_{\mathcal E,\mathcal E'}
    C_{r;\mathcal E}C_{r;\mathcal E'}^{*}
    &\times
    \operatorname{Tr}_{\overline{S_1}}
    \left[
        W_{\mathcal E}
        (\rho_A\otimes\Omega)
        W_{\mathcal E'}^\dagger
    \right].
\end{aligned}
\label{eq:correlated_general_output}
\end{equation}

Equation(\ref{eq:correlated_general_output}) highlights the essential difference from the local single-qubit cases. In Secs.\ref{sec:auxiliary_noise} and \ref{sec:core qubits}, orthogonality of the Bell-pair factors eliminates the cross terms between distinct local Pauli components, so that the reduced recovery channel depends only on the diagonal Pauli weights $q_a$. For a general correlated channel, distinct multiqubit Pauli strings $P_{\mathcal{E}}$ and $P_{\mathcal{E}'}$ need not become orthogonal on the subsystems that are traced out. Consequently, terms with $\mathcal{E}\neq\mathcal{E}'$ may survive, and the effective recovery channel can depend on the full multiqubit process matrix rather than only on a set of Pauli probabilities.

A useful special case is a correlated Pauli channel whose process matrix is already diagonal in the multiqubit Pauli basis. In that case, Eq.(\ref{eq:correlated_general_output}) reduces to a classical mixture over the propagated Pauli strings $W_{\mathcal{E}}$. Different physical Pauli strings may still become indistinguishable after the complementary subsystems are traced out, but no coherence between distinct strings needs to be retained.

The results of Sec.\ref{sec:general_channels} therefore establish a hierarchy of effective noise propagation. Noise on an auxiliary signal qubit is invisible to the reduced recovered state; local noise on an auxiliary key qubit becomes a Pauli-diagonal effective channel; core-qubit noise additionally exhibits parity-dependent permutations or merging of Pauli sectors; and general correlated noise on the decryption subsystem may require the full multiqubit process matrix. We next specialize these results to depolarizing noise, for which the effective recovery channels and fidelities can be obtained explicitly.

\section{Depolarizing Noise}
\label{sec:depolarizing noise}

\subsection{Local depolarizing noise}
\label{sec:Local depolarizing noise}
We now specialize the general results of Sec.\ref{sec:general_channels} to local depolarizing noise. 
For a single qubit, we define the depolarizing channel as
\begin{equation}
\mathcal{D}_p(\rho)=p\rho+(1-p)\frac{I}{2},\qquad
0\le p\le 1.
\label{eq:depolarizing}
\end{equation}
Equivalently, it can be written in Pauli form as
\begin{equation}
\mathcal{D}_p(\rho)=q_0\rho+\sum_{a=1}^{3}
q_a\sigma_a\rho\sigma_a,
\label{eq:depolarizing_pauli}
\end{equation}
with
\begin{equation}
q_0=\frac{1+3p}{4},
\qquad
q_1=q_2=q_3=\frac{1-p}{4}.
\label{eq:depolarizing_weights}
\end{equation}
Because the three nontrivial Pauli components have equal weights, any parity dependence that amounts only to a permutation of $\sigma_1,\sigma_2,\sigma_3$ becomes irrelevant.

For an auxiliary signal qubit $S_j$, $j\ge2$, the result of Sec.\ref{sec:auxiliary_noise} applies independently of the depolarizing parameter:
\[\rho_{S_1}^{\rm out}=
\rho_A^{(S_1)}.
\]
Thus, local depolarization of $S_j$ has no effect on the reduced recovered state.

For an auxiliary key qubit $N_j$, $j\ge2$, substituting Eq.~\eqref{eq:depolarizing_weights} into Eq.(\ref{eq:Nj_effective_pauli}) gives
\[\rho_{S_1}^{\rm out}=p_{N_j}\rho_A^{(S_1)}+
(1-p_{N_j})\frac{I_{S_1}}{2}.
\]
Hence, the physical depolarizing channel is transferred to the recovered qubit as a depolarizing effective channel with the same parameter.

For the core signal qubit $S_1$, Eqs.(\ref{eq:S1_even}) and (\ref{eq:S1_odd}) differ only by a permutation of the three nontrivial Pauli sectors. 
Since the depolarizing channel satisfies $q_1=q_2=q_3$, this permutation has no effect, and for either parity one obtains
\[
\rho_{S_1}^{\rm out}=p_{S_1}\rho_A^{(S_1)}+(1-p_{S_1})\frac{I_{S_1}}{2}.
\]
Thus, local depolarization of $S_1$ remains isotropic in the effective recovery channel.

The core key qubit $N_1$ behaves differently. 
For even $n$, Eq.(\ref{eq:N1_even}) again involves only a permutation of the nontrivial Pauli sectors, yielding
\[
\rho_{S_1}^{\rm out}=p_{N_1}\rho_A^{(S_1)}+(1-p_{N_1})\frac{I_{S_1}}{2}.
\]
For odd $n$, however, the Pauli-sector merging in Eq.(\ref{eq:N1_odd}) gives
\[q_0+q_2=\frac{1+p_{N_1}}{2},\qquad q_1+q_3
=
\frac{1-p_{N_1}}{2},
\]
and therefore
\begin{equation}
\rho_{S_1}^{\rm out}=\frac{1+p_{N_1}}{2}\rho_A^{(S_1)}+\frac{1-p_{N_1}}{2}\sigma_2^{(S_1)}\rho_A^{(S_1)}\sigma_2^{(S_1)}.
\label{eq:N1_odd_depolarizing}
\end{equation}
Thus, an isotropic physical depolarizing channel on $N_1$ is converted into an anisotropic effective Pauli channel when $n$ is odd.

Equation~\eqref{eq:N1_odd_depolarizing} coincides with the odd-$n$ result obtained in Sec.\ref{subsec:Werner resource} for a Werner-type imperfection of the core pair, upon identifying $p_{N_1}=p_1$. 
This agreement follows from the channel--resource correspondence discussed in Sec.\ref{subsec:Werner resource to channel noise}, since $N_1$ does not participate in the encryption unitary.

Local depolarizing noise therefore exhibits three distinct behaviors. 
Noise on $S_j$ ($j\ge2$) is invisible to the reduced recovered state; noise on $N_j$ ($j\ge2$), on $S_1$, and on $N_1$ for even $n$ produces an ordinary depolarizing effective channel with the same parameter; only noise on $N_1$ for odd $n$ is converted into an anisotropic two-sector Pauli channel.

\subsection{Distributed versus global depolarization}
\label{sec:Distributed versus global depolarization}
We next compare two physically distinct multiqubit depolarizing-noise models acting on the decryption subsystem. 
Depolarization provides a useful standard description of noisy multiqubit quantum operations\cite{dur2005standard}. 
We first consider a distributed product model in which the core pair $S_1N_1$ undergoes a two-qubit depolarizing channel, while a subset of the auxiliary key qubits independently undergo local single-qubit depolarizing channels.

Let
\[B_s=S_1N_1N_2\cdots N_s,\qquad1\le s\le n.
\]
The distributed product channel is defined as
\begin{equation}
\Lambda_{B_s}^{\rm prod}=\mathcal{D}_{p_1}^{(S_1N_1)}\otimes\bigotimes_{j=2}^{s}
\mathcal{D}_{p_j}^{(N_j)} .
\label{eq:product_depolarization}
\end{equation}

Here the first factor is a single two-qubit depolarizing channel acting collectively on the core pair $S_1N_1$; it should not be confused with independent single-qubit depolarization of $S_1$ and $N_1$. 
The two-qubit depolarizing channel is defined by
\begin{equation}
\mathcal{D}_{p_1}^{(S_1N_1)}(\rho)=p_1\rho+(1-p_1)\frac{I_{S_1N_1}}{4}
\otimes\operatorname{Tr}_{S_1N_1}(\rho).
\label{eq:core_pair_depolarization}
\end{equation}
Consequently, the odd-$n$ anisotropy associated with depolarization of $N_1$ alone, discussed in Sec.\ref{sec:Local depolarizing noise}, does not apply to the collective core-pair channel considered here.

The auxiliary key-qubit channels in Eq.(\ref{eq:product_depolarization}) admit a useful alternative interpretation. As discussed in Sec.\ref{subsec:Werner resource to channel noise}, the key qubits $N_j$ do not participate in the encryption operation. 
Therefore, for $j\ge2$,
\[\mathcal{D}_{p_j}^{(N_j)}\circ{U}_{\rm enc}^{(n)}={U}_{\rm enc}^{(n)}
\circ\mathcal{D}_{p_j}^{(N_j)} .
\]
Each post-encryption depolarizing channel on $N_j$ may therefore be shifted to the resource-preparation stage. 
Acting on one half of the initially ideal Bell pair gives
\[
\left(\mathcal{I}_{S_j}\otimes\mathcal{D}_{p_j}^{(N_j)}
\right)
\left(\Phi^+_{S_jN_j}\right)
=
p_j\Phi^+_{S_jN_j}+(1-p_j)\frac{I_{S_jN_j}}{4}\equiv
W_{p_j}^{(S_jN_j)} .
\]

Hence, the distributed model in Eq.~\eqref{eq:product_depolarization} is equivalently described as follows: the auxiliary pairs $S_jN_j$, $2\le j\le s$, are initially prepared in Werner-type states with parameters $p_j$, while the core pair $S_1N_1$ undergoes the two-qubit depolarizing channel $\mathcal{D}_{p_1}^{(S_1N_1)}$ after encryption. 
This equivalence applies to the auxiliary key-qubit channels because they commute with the encryption operation; the core-pair channel cannot in general be shifted in the same way because $S_1$ participates nontrivially in $U_{\rm enc}^{(n)}$.

With the encryption and decryption,we obtain corresponding reduced recovered state is
\begin{equation}
\rho_{S_1}^{\rm out}=p_{\rm eff}^{\rm prod}\rho_A^{(S_1)}+\left(
1-p_{\rm eff}^{\rm prod}\right)\frac{I_{S_1}}{2}.
\label{eq:product_recovered_state}
\end{equation}

where $p_{\rm eff}^{\rm prod}=\prod_{j=1}^{s} p_j$. For the uniformly distributed case, we take $s=n$ and set
\begin{equation}
p_1=p_2=\cdots=p_n=p .
\label{eq:uniform_depolarization}
\end{equation}
The effective parameter then becomes
\begin{equation}
p_{\rm eff}^{\rm prod}=p^n ,
\label{eq:uniform_effective_parameter}
\end{equation}
so that
\begin{equation}
\rho_{S_1}^{\rm out}=p^n\rho_A^{(S_1)}+
\left(1-p^n\right)
\frac{I_{S_1}}{2}.
\label{eq:uniform_recovered_state}
\end{equation}

We now compare this distributed product model with a genuine global depolarizing channel acting on the entire decryption subsystem
\begin{equation}
B=S_1N_1N_2\cdots N_n,
\qquad
d_B=2^{n+1}.
\label{eq:decryption_subsystem}
\end{equation}
The global depolarizing channel is defined by
\begin{equation}
\mathcal{D}_{P}^{(B)}(\rho)=P\rho+(1-P)
\frac{I_B}{2^{n+1}}
\otimes
\operatorname{Tr}_{B}(\rho).
\label{eq:global_depolarization}
\end{equation}

The two branches of this channel have a direct interpretation after decryption. 
The first branch, with weight $P$, leaves the encrypted state unchanged and therefore gives the ideal recovered state $\rho_A^{(S_1)}$. 
For the second branch, the subsystem $B$ is replaced by the maximally mixed state. 
Since the maximally mixed state is invariant under the decryption unitary,
\[
U_{\rm dec}^{(n)}
\frac{I_B}{2^{n+1}}
U_{\rm dec}^{(n)\dagger}
=
\frac{I_B}{2^{n+1}},
\]
its reduced state on $S_1$ is $I_{S_1}/2$. 
Consequently,
\begin{equation}
\rho_{S_1}^{\rm out}
=
P\rho_A^{(S_1)}
+
(1-P)\frac{I_{S_1}}{2}.
\label{eq:global_recovered_state}
\end{equation}

The distributed product model and the global depolarizing model are physically distinct. 
In particular,
\begin{equation}
\mathcal{D}_{p}^{(S_1N_1)}
\otimes
\bigotimes_{j=2}^{n}
\mathcal{D}_{p}^{(N_j)}
\neq
\mathcal{D}_{p}^{(B)}
\label{eq:product_not_global}
\end{equation}
in general. 
The corresponding effective depolarizing parameters observed on the recovered qubit are
\begin{equation}
p_{\rm eff}^{\rm prod}=p^n,
\qquad
p_{\rm eff}^{\rm global}=P .
\label{eq:effective_parameter_comparison}
\end{equation}
Nevertheless, the two physically different noise models induce the same effective recovery channel on $S_1$ when
\begin{equation}
P=p^n .
\label{eq:matching_condition}
\end{equation}

This equality applies only to the reduced effective channel observed on the recovered qubit. 
It does not imply equality of the underlying multiqubit physical channels. 
The product and global models act differently on the full decryption subsystem and can therefore generate different global states and correlations even when their reduced recovery channels on $S_1$ coincide.

\subsection{Recovery fidelity}

Having obtained the effective recovery channels for the different depolarizing-noise configurations, we now quantify the quality of recovery in terms of the state fidelity. 
For a pure input state
\begin{equation}
\rho_A=|\psi\rangle\langle\psi|_A,
\label{eq:pure_input}
\end{equation}
the recovery fidelity is defined as\cite{jozsa1994fidelity}
\begin{equation}
F=\langle\psi|\rho_{S_1}^{\rm out}
|\psi\rangle .
\label{eq:recovery_fidelity}
\end{equation}

For the isotropic effective channels considered above, the recovered state takes the form
\begin{equation}
\rho_{S_1}^{\rm out}=p_{\rm eff}\rho_A^{(S_1)}+(1-p_{\rm eff})\frac{I_{S_1}}{2}.
\label{eq:isotropic_effective_state}
\end{equation}
Substituting Eq.~\eqref{eq:isotropic_effective_state} into Eq.~\eqref{eq:recovery_fidelity} gives
\begin{equation}
F=\frac{1+p_{\rm eff}}{2}.
\label{eq:isotropic_fidelity}
\end{equation}
Thus, whenever the effective recovery channel remains isotropic, the fidelity is independent of the input-state direction and is determined entirely by the effective depolarizing parameter.

The location dependence derived in Secs.\ref{sec:general_channels} and \ref{sec:Local depolarizing noise} is directly reflected in the recovery fidelity. 
For depolarizing noise acting on an auxiliary signal qubit $S_j$, $j\ge2$, the reduced recovered state is unchanged, and therefore
\begin{equation}
F_{S_j}=1,
\qquad j\ge2.
\label{eq:fidelity_Sj}
\end{equation}
For an auxiliary key qubit $N_j$, $j\ge2$, the physical depolarizing parameter is transferred directly to the effective recovery channel, yielding
\begin{equation}
F_{N_j}
=
\frac{1+p_j}{2},
\qquad j\ge2.
\label{eq:fidelity_Nj}
\end{equation}
Likewise, depolarization of the core signal qubit $S_1$ gives
\begin{equation}
F_{S_1}
=
\frac{1+p_{S_1}}{2}
\label{eq:fidelity_S1}
\end{equation}
for both even and odd $n$. 
The same expression holds for depolarization of the core key qubit $N_1$ when $n$ is even:
\begin{equation}
F_{N_1}^{\rm even}
=
\frac{1+p_{N_1}}{2}.
\label{eq:fidelity_N1_even}
\end{equation}

A qualitatively different behavior occurs for depolarizing noise acting on $N_1$ when $n$ is odd. 
In this case, the Pauli-sector merging derived in Sec.\ref{sec:core qubits} produces the anisotropic recovered state
\begin{equation}
\rho_{S_1}^{\rm out}
=
\frac{1+p_{N_1}}{2}\rho_A^{(S_1)}
+
\frac{1-p_{N_1}}{2}
\sigma_2^{(S_1)}
\rho_A^{(S_1)}
\sigma_2^{(S_1)}.
\label{eq:N1_odd_state_fidelity_section}
\end{equation}
The corresponding fidelity is
\begin{equation}
F_{N_1}^{\rm odd}
=
\frac{1+p_{N_1}}{2}
+
\frac{1-p_{N_1}}{2}
\left|
\langle\psi|\sigma_2|\psi\rangle
\right|^2.
\label{eq:N1_odd_fidelity}
\end{equation}
Unlike the isotropic result in Eq.~\eqref{eq:isotropic_fidelity}, this fidelity depends explicitly on the direction of the input state on the Bloch sphere.

The minimum is obtained for states satisfying
$\langle\psi|\sigma_2|\psi\rangle=0$, for which
\begin{equation}
F_{\min}
=
\frac{1+p_{N_1}}{2}.
\label{eq:fidelity_min}
\end{equation}
By contrast, an eigenstate of $\sigma_2$ satisfies
\begin{equation}
\left|
\langle\psi|\sigma_2|\psi\rangle
\right|^2
=
1,
\label{eq:sigma2_eigenstate}
\end{equation}
and is therefore recovered perfectly:
\begin{equation}
F_{\max}=1.
\label{eq:fidelity_max}
\end{equation}

For pure input states uniformly distributed over the Bloch sphere, the state average satisfies
\begin{equation}
\overline{
\left|
\langle\psi|\sigma_2|\psi\rangle
\right|^2
}
=
\frac{1}{3}.
\label{eq:bloch_average}
\end{equation}
Hence, the Bloch-sphere-averaged recovery fidelity is
\begin{equation}
\overline{F}_{N_1}^{\rm odd}
=
\frac{2+p_{N_1}}{3}.
\label{eq:N1_odd_average_fidelity}
\end{equation}

We next consider the distributed product-depolarization model introduced in Sec.\ref{sec:Distributed versus global depolarization}. 
For parameters $p_1,p_2,\ldots,p_s$, the effective depolarizing parameter is
\begin{equation}
p_{\rm eff}^{\rm prod}
=
\prod_{j=1}^{s}p_j ,
\label{eq:product_parameter_fidelity}
\end{equation}
and the corresponding fidelity is
\begin{equation}
F_{\rm prod}
=
\frac{
1+\displaystyle\prod_{j=1}^{s}p_j
}{2}.
\label{eq:product_fidelity}
\end{equation}
For the uniformly distributed case considered in Sec.\ref{sec:Distributed versus global depolarization}, where the full decryption subsystem is involved ($s=n$) and
$p_1=p_2=\cdots=p_n=p$, one has
\begin{equation}
p_{\rm eff}^{\rm prod}
=
p^n,
\label{eq:uniform_product_parameter_fidelity}
\end{equation}
and therefore
\begin{equation}
F_{\rm prod}(p,n)
=
\frac{1+p^n}{2}.
\label{eq:uniform_product_fidelity}
\end{equation}
For any fixed $0\le p<1$, the fidelity decreases monotonically with $n$ and approaches $1/2$ as $n\rightarrow\infty$. 
This behavior reflects the decreasing weight of the branch in which all components of the distributed model preserve the ideal recovery.

By contrast, consider the genuine global depolarizing channel acting on
\[
B=S_1N_1N_2\cdots N_n .
\]
As obtained in Sec.\ref{sec:Distributed versus global depolarization}, the recovered state is
\begin{equation}
\rho_{S_1}^{\rm out}
=
P\rho_A^{(S_1)}
+
(1-P)\frac{I_{S_1}}{2},
\label{eq:global_state_fidelity_section}
\end{equation}
which gives
\begin{equation}
F_{\rm global}
=
\frac{1+P}{2}.
\label{eq:global_fidelity}
\end{equation}
Unlike the uniformly distributed product model, the global channel is characterized by a single collective parameter $P$ and therefore does not generate an additional power of $n$. 
When $P=p^n$, the two models induce the same effective recovery channel and hence the same recovery fidelity. 
As emphasized in Sec.\ref{sec:Distributed versus global depolarization}, this equality does not imply that the underlying multiqubit physical channels are identical.

\begin{figure*}[t]
    \centering
    \includegraphics[width=1.0\textwidth]
    {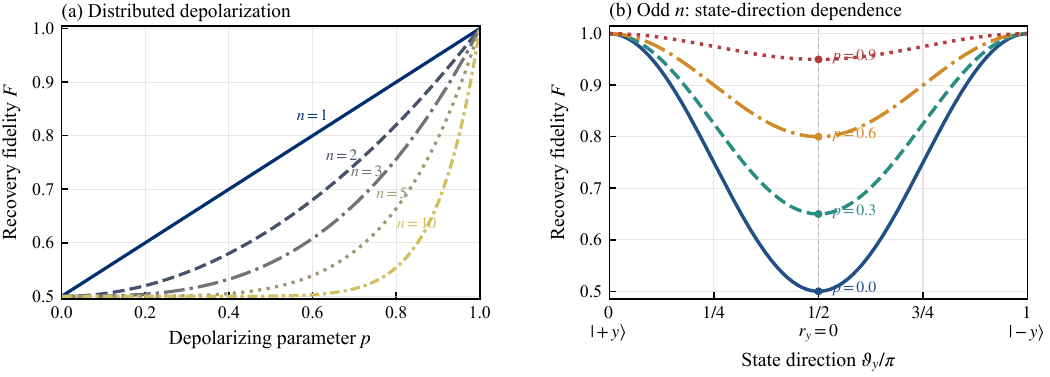}
    \caption{
    Recovery fidelity under depolarizing noise.
(a) Fidelity for the uniformly distributed product-depolarization model of Sec.\ref{sec:Distributed versus global depolarization}, in which the core pair $S_1N_1$ undergoes a two-qubit depolarizing channel and the auxiliary key qubits $N_2,\ldots,N_n$ undergo independent single-qubit depolarizing channels, all with the same parameter $p$. The effective parameter is $p_{\rm eff}^{\rm prod}=p^n$, giving
$F_{\rm prod}(p,n)=(1+p^n)/2$.
The curves correspond to $n=1,2,3,5,$ and $10$; $n=1$ corresponds to the core-pair channel alone and is included as a reference.
(b) Input-state dependence produced by depolarization of the core key qubit $N_1$ for odd $n$.
Writing
$\cos\vartheta_y=\langle\psi|\sigma_y|\psi\rangle=r_y$,
the fidelity is
$F_{N_1}^{\rm odd}(p,\vartheta_y)
=(1+p)/2+(1-p)\cos^2\vartheta_y/2$.
The curves correspond to $p=0,0.3,0.6,$ and $0.9$.
The states $|\pm y\rangle$ are recovered perfectly, whereas states with $r_y=0$ attain the minimum fidelity
$F_{\min}=(1+p)/2$.
    }
    \label{fig:fidelity_curves}
\end{figure*}

The two characteristic fidelity behaviors are illustrated in Fig.\ref{fig:fidelity_curves}. 
Figure\ref{fig:fidelity_curves}(a) shows the multiplicative reduction of the effective parameter in the uniformly distributed product-depolarization model, while Fig.\ref{fig:fidelity_curves}(b) shows the input-state dependence produced by depolarization of the core key qubit $N_1$ for odd $n$. 
The latter directly reflects the anisotropic Bloch-vector transformation derived in Sec.\ref{sec:Local depolarizing noise}: the $y$ component is preserved, whereas the $x$ and $z$ components are reduced by the physical depolarizing parameter.

These results demonstrate that recovery fidelity is controlled not only by the strength of the physical noise, but also by its location, the parity-dependent structure of error propagation, and whether the underlying multiqubit noise is distributed or global.

\section{Discussion And Conclusion}
In this work, we have developed a channel-based framework for describing how physical noise propagates through the encryption--decryption structure of encrypted quantum cloning. 
Rather than characterizing imperfections only through the final recovered state, we formulated the problem in terms of an effective recovery channel
\[
\mathcal{E}_{X}^{\rm eff}:
\mathcal{B}(\mathcal{H}_A)
\rightarrow
\mathcal{B}(\mathcal{H}_{S_1}),
\]
which maps the original input state on $A$ to the recovered state on $S_1$ when a physical channel acts at location $X$. 
By expanding local Kraus operators in the Pauli basis and propagating the corresponding Pauli components through the decryption unitary, the influence of the physical noise can be characterized directly in terms of its location in the encrypted-cloning architecture.

The resulting effective noise exhibits a clear hierarchy. 
A trace-preserving channel acting only on an auxiliary signal qubit $S_j$ ($j\ge2$) leaves the reduced recovered state on $S_1$ unchanged, because $S_j$ does not participate in the decryption unitary and is subsequently traced out. 
In contrast, an arbitrary local single-qubit CPTP channel acting on an auxiliary key qubit $N_j$ ($j\ge2$) is converted into a Pauli-diagonal effective channel on $S_1$. 
Its effective Pauli probabilities are given by the Bell-basis populations of the Choi state of the physical channel, showing that only the corresponding Bell-diagonal information is retained in the reduced recovery channel. 
For the core qubits, the propagation acquires an additional dependence on the parity of the number of entangled pairs. 
Noise on $S_1$ leads only to a parity-dependent permutation of the nontrivial Pauli sectors, whereas noise on $N_1$ exhibits a stronger effect: for odd $n$, distinct physical Pauli components merge into the same effective sector. 
In particular,
\[
\{I,\sigma_2\}
\longrightarrow I_{\rm eff},
\qquad
\{\sigma_1,\sigma_3\}
\longrightarrow \sigma_{2,\rm eff}.
\]
Thus, the decryption procedure acts not only as a state-recovery transformation, but also as a structured map that can erase, permute, or merge information about physical error sectors.

Depolarizing noise makes these structural effects especially transparent. 
For noise on $S_1$ or on $N_1$ with even $n$, as well as on an auxiliary key qubit $N_j$, the effective recovery channel remains isotropic. 
By contrast, for depolarizing noise acting on $N_1$ with odd $n$, the isotropic physical channel is converted into
\[
\mathcal{E}_{N_1}^{\rm eff}(\rho)
=
\frac{1+p}{2}\rho
+
\frac{1-p}{2}\sigma_2\rho\sigma_2,
\]
The physical isotropy is therefore lost at the level of the effective recovery channel, while the $y$ direction remains protected. 
This explains both the input-state dependence of the recovery fidelity and the perfect recovery of the $\sigma_2$ eigenstates. 
The depolarizing model also provides a direct connection between dynamical noise and imperfect entanglement resources: local depolarization of one half of an ideal Bell pair produces the Werner-type resource considered in Sec.\ref{sec:werner}, in accordance with the broader channel--state and teleportation-based connections discussed in Refs\cite{bowen2001teleportation,ziman2005concurrence,pirandola2017fundamental}. 
Importantly, this correspondence depends on the location of the physical channel. 
Depolarization on a key qubit $N_j$ can be shifted through the encryption operation because $N_j$ is not acted upon by $U_{\rm enc}^{(n)}$, whereas a channel acting on a subsystem involved nontrivially in encryption cannot in general be interpreted in this way.

The same distinction is important for multiqubit depolarization. 
In the distributed product model considered in Sec.\ref{sec:Distributed versus global depolarization}, the auxiliary key-qubit channels may be shifted to the resource-preparation stage and interpreted as Werner-type auxiliary resources, while the core pair undergoes a genuine post-encryption two-qubit depolarizing channel. 
The resulting effective parameter is
\[
p_{\rm eff}^{\rm prod}
=
\prod_{j=1}^{s}p_j,
\]
and becomes $p^n$ in the uniform full-subsystem case. 
A genuine global depolarizing channel, by contrast, is characterized by a single collective parameter $P$. 
Although the two physically different models induce the same effective recovery channel when $P=p^n$, this equality holds only at the level of the recovered qubit and does not imply equality of the underlying multiqubit channels, global states, or correlations. 
This distinction illustrates that the effective channel observed after recovery does not, in general, uniquely identify the microscopic noise process that produced it.

Our treatment of correlated noise further shows the limits of the single-qubit Pauli-probability description. 
For a general correlated CPTP channel acting on the decryption subsystem, coherences between different multiqubit Pauli strings may survive the final partial trace, so that the full multiqubit process matrix can be required. 
Correlated Pauli channels form an important simpler subclass, because their process matrices are already diagonal in the multiqubit Pauli basis. 
A more systematic investigation of correlated noise and quantum channels with memory, including time-correlated amplitude-damping models, would therefore be a natural extension of the present analysis\cite{macchiavello2002entanglement,yeo2003quantum,daffer2004depolarizing,kretschmann2005quantum,
caruso2010teleportation,RevModPhys.86.1203,yeo2003time}. 
The bit-flip and generalized amplitude-damping examples considered here also indicate that the framework applies beyond isotropic unital noise, including anisotropic Pauli noise and nonunital dynamics\cite{nielsen2010quantum,srikanth2008squeezed}. 
It would be useful to classify more generally which properties of a physical channel---such as unitality, anisotropy, and preferred-axis structure---are preserved, erased, or transformed by the encrypted-cloning recovery procedure. 
Imperfections in the encryption and decryption operations themselves provide another natural direction. 
Finally, a systematic channel-level comparison using entanglement fidelity, average gate fidelity, and channel-distance measures could complement the state-recovery fidelities considered here and provide a more complete characterization of the resulting effective channels\cite{schumacher1996sending,nielsen2002simple,gilchrist2005distance,fuchs1999cryptographic}.

Overall, our results show that the effect of physical noise in encrypted quantum cloning is not determined by noise strength alone. 
The subsystem on which the channel acts, the parity of the protocol, and correlations among noisy subsystems jointly determine the effective channel experienced by the recovered qubit. 
In this sense, the encrypted-cloning decryption structure acts as a nontrivial noise-transduction mechanism: physically different errors may be removed from the recovered subsystem, transferred as Pauli errors, permuted, or merged into the same effective error sector.

\begin{acknowledgments}
This work was supported by the National Natural Science Foundation of China (Grant Nos.12464064 and 12165020), the Major Science and Technology Project of Yunnan Province (Grant No.202502AD080015), and the Yunnan Fundamental Research Project (Grant No.202401AT070448). L.-J. Wang acknowledges support from the Xing Dian Talent Program (Grant No.C619300A090). 
\end{acknowledgments}

\bibliography{reference}
\bibliographystyle{ieeetr}


\end{document}